\documentclass{svproc}
\usepackage{url}
\usepackage{dirtytalk}
\usepackage{graphicx}
\usepackage{multirow}

\usepackage{sidecap}

\usepackage{booktabs}
\usepackage{color, colortbl}
\definecolor{LightGrey}{rgb}{0.9,0.9,0.9}

\begin{document}
\mainmatter              
\title{Beyond Fortune 500: Women in a Global Network of Directors}
\titlerunning{Women on boards}  
%
\author{Anna Evtushenko\inst{1} \and Michael T.~Gastner\inst{2}}
\authorrunning{Evtushenko and Gastner} 
%
\tocauthor{Anna Evtushenko and Michael T.~Gastner}
\institute{Cornell University, Ithaca, NY 14853, USA,\\
\email{ae392@cornell.edu},\\
\and
Yale-NUS College,
16 College Avenue West, \#01-220 Singapore 138527, Singapore}

\maketitle              

\begin{abstract}
In many countries, the representation of women on corporate boards of directors has become a topic of intense political debate.
Social networking plays a crucial role in the appointment to a board so that an informed debate requires knowing where women are located in the network of directors.
One way to quantify the network is by studying the links created by serving on the same board and by joint appointments on multiple boards.
We analyse a network of $\approx 320\,000$ board members of $36\,000$ companies traded on stock exchanges all over the world, focusing specifically on the position of women in the network. 
Women only have $\approx 9-13\%$ of all seats, but they are not marginalised.
Applying metrics from social network analysis, we find that their influence is close to that of men.  
We do not find evidence to support previous claims that women play the role of ``queen bees'' that exclude other women from similar positions.
\end{abstract}

\section{Introduction}
Females on boards of directors and board diversity more broadly are the topic of many studies~\cite{adams2016women,delis2016effect,gabaldon2016searching}.
Research has shown that female board representation is ``positively related to accounting returns'' \cite{post2015women}.
The World Bank~\cite{WorldBank17} estimates that 39\% of the
worldwide labour force in 2016 are women, but the percentage of women
in leadership positions is much lower. 
Recent reports state that only 24\% of senior management
positions~\cite{GrantThornton16} and 15\% of corporate board
seats~\cite{Dawson_etal16} are held by women.
The percentage of female CEOs among Fortune 500 firms is even lower (6.4\%)~\cite{mcgregor_2017}.
Women's chances to become a CEO or a board member depend on multiple factors, such as ``country wealth, gender egalitarianism and humane orientation'' \cite{elango2018women}.
Nevertheless, female board participation is slowly on the
rise globally.
Shareholders and governments no longer regard it as a legitimate practice to recruit directors from an exclusively male ``old-boys network''~\cite{Perrault15,dHoop_etal17}.
Various countries, for example Israel~\cite{Izraeli00} and Norway~\cite{SeierstadOpsahl11}, have enacted laws that favour the appointment of women~\cite{sojo2016reporting}. 
As a consequence, there are signs that, at least in some European countries, the \say{glass ceiling} that has kept women out of the
boardrooms is beginning to crack~\cite{EuropeanCommission15}.

Once appointed, female directors must navigate intricate networks
of professional relationships.
A concrete manifestation of such a professional network are \say{interlocking directorates}~\cite{LevineRoy79} (i.e.\ the practice that some directors hold seats on more than one board).
A recent survey by Credit Suisse~\cite{Dawson_etal16b} relates
\say{overboarding} (i.e.\ an excessive number of board seats held by an
individual) to the current trend towards increasing the number of
female board members. The probability that a woman joins the board has been shown to be negatively correlated with the number of women currently on the board and to increase when a woman departs the board \cite{farrell2005additions}.
The underlying assumption is that companies tend to recruit \say{token
women} (i.e.\ exactly one per board) from a limited pool of female candidates~\cite{dezsHo2016there,StrydomYong12}. Some commentators compare women directors with multiple seats to queen bees~\cite{ThomsonGraham05}, implying that these women allegedly usurp power at the expense of female competitors. The purpose of this article is to test whether such narratives stand up to quantitative scrutiny.

\begin{figure}
    \centering
    \hspace*{-1cm}
    \includegraphics[width=\textwidth]{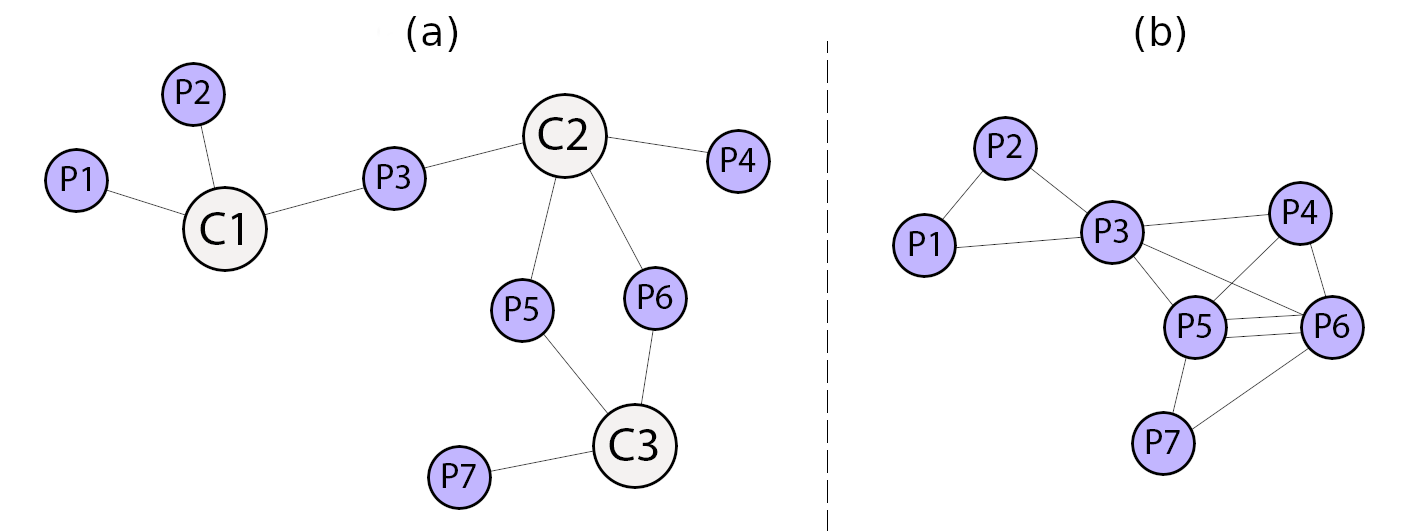}
    \caption{Network representations of board interlock.
    (a) Bipartite graph where every edge is between a company (large node) and a person (i.e.\ a director, small node).
    (b) In this article, we analyse the social network of directors that results from a one-mode projection of the bipartite graph:
    directors are connected if and only if they sit together on a board.}
    \label{fig:netw}
\end{figure}

Board interlocks can be inferred from a bipartite graph where every
edge is between one company and one director (Figure~\ref{fig:netw})~\cite{BattistonCatanzaro04}.
Each director at one end of an edge sits on the board of the company at
the other end of this edge.
The study of board interlock networks started already in the
1970s~\cite{SonquistKoenig75}, but at that time the role of the
director's gender was not yet in the limelight.
Interest in the role of women on boards has intensified in recent years, see for example Ref.~\cite{Eagly16} for a critical review.
Still, relatively little is known about the role that the gender plays for
the network formed by interlocking directorates.

A woman's position in the social network certainly influences her chances to be appointed to a board~\cite{BurgessTharenou02}, but only few papers have analysed the real network~\cite{WestphalMilton00,Hillman_etal07,SeierstadOpsahl11,Martinez12}.
The most comprehensive quantitative network study to date is the PhD thesis by Hawarden~\cite{Hawarden10}, which looks at empirical data and builds a modelling framework called \say{Glass Network} theory.
It posits the existence of ``glass nets'' that prevent women from assuming board seats, but those women who succeeded in crossing a glass net behave like queen bees.
We further the exploration of this topic using network analysis applied to a large dataset, which we now describe.

\section{Data}
\label{sec:data}

The source of our data is the Financial Times~\cite{ThomsonReuters16}, an international English-language newspaper specialising in business and economics. Its website is a source of up-to-date information on financial markets and companies traded as equities. The website has data on the performance and structure of roughly 36\,000 companies from 54 countries. The resulting data base is, to the best of our knowledge, the largest that has so far been used in the literature on board interlocks. The Financial Times (FT) receives their information from the media company Thomson Reuters, which in principle has data on yet more firms. However, we decided to use the FT data because this subset is more representative of companies that make up the business world.  

The specific fields that we obtained for each company were name, unique code, sector, industry (i.e.\ subsector), country, revenue for the past 12 months, number of employees, date incorporated, and a list of directors, each if available. For each director, we recorded his or her name, gender, and age, each if included in the FT database. People were then identified as the same individual and assigned a unique ID if their three fields matched (for example, the names and genders were the same, and ages were blank) because we assume that the underlying Thomson Reuters database would have identical entries on each individual in all his or her companies. There were a total of 35\,927 companies and 321\,967 directors. Here we make no distinction between executive and non-executive directors.

In terms of missing data, 273 companies have zero directors listed. Among the fields relevant to the analysis, for 5732 companies (15.95\%) we have no information about the country, for 4461 (12.41\%) no sector, and for 5020 (13.97\%) no industry. 96\,751 directors (30.1\%) are listed without gender. For 126\,092 directors (39.2\%), the database contains no information about their age.
We show summary statistics of the network and the subgraphs consisting of only male or only female nodes in Table~\ref{table:subgraphs}.

\begin{center}
\begin{table}
\renewcommand{\arraystretch}{1.2}
\renewcommand{\aboverulesep}{0pt}
\renewcommand{\belowrulesep}{0pt}
\centering
  \caption{Statistics of the full network and the subgraphs consisting of only male or only female nodes. We calculate the average path length with the harmonic mean formula by Newman~\cite{Newman10} to handle disconnected components.}
  \label{table:subgraphs}
  \begin{minipage}{\textwidth}
    \begin{tabular}{p{0.45\textwidth}p{0.17\textwidth}p{0.17\textwidth}p{0.16\textwidth}}
      \toprule
      \rowcolor{LightGrey}
    \textbf{nodes} & \textbf{all} & \textbf{male} & \textbf{female}\\
      \midrule
edges & 2809623 & 1092004 & 44666 \\ 
\hline
diameter & 24 & 29 & 40 \\ 
\hline
average path length & 13.90 & 22.90 & 517.79 \\ 
\hline
density & $5.4 \times 10^{-5}$ & $5.7\times 10^{-5}$ & $9.6\times 10^{-5}$ \\
\hline
components & 9404 & 12393 & 12355 \\ 
\hline
\% of nodes in the largest component & 74.7 & 74.8 & 79.4 \\ 
\hline
\% of edges in the largest component & 85.5 & 85.1 & 89.6 \\ 
      \bottomrule
    \end{tabular}
    \vspace{-2\baselineskip}
  \end{minipage}
  \label{table_sub}
\end{table}
\end{center}

\section{Summary Statistics of Node Attributes}

The proportion of female directors in the FT data is 9.43\% of all nodes and 13.49\% among people with confirmed gender.
This percentage is comparable to values stated in previous studies of international data.
For example, Dawson et al.~\cite{Dawson_etal16} found that women hold 14.7\% of seats in the CS Gender 3000 data base; Deloitte~\cite{Deloitte17} puts this number at 15\% for data from nearly 7000 companies.
Both of these reports emphasise that there can conceivably be a difference between the proportion of female {\it directors} and the proportion of female {\it seats} because of overboarding.
On the boards of S\&P500 companies, overboarding is more prevalent among women~\cite{Dawson_etal16b}. 
However, in the more comprehensive FT data, we find that, at the international level, the proportions of seats and directors are similar: the percentage of female seats is 9.73\% and thus only 0.29\% larger than the percentage of female directors.
The underlying reason is that overboarding hardly differs between genders: 14.7\% of women and 14.8\% of men are multiple directors. Overall, taking into account ungendered nodes, 14.06\% of directors are multiple directors. 4.22\% of directors are on more than 2 boards, 1.76\% on more than 3, and 0.40\% on more than 5. 

It is interesting to see whether the ratio of female seats to all seats differs by country, sector or industry. We can easily compute these numbers because for each company we know its country, sector, and industry, as well as the number of all directors and the number of female directors.

\begin{figure}
\centering
\hspace*{-0.5cm}
\includegraphics[height=14cm]{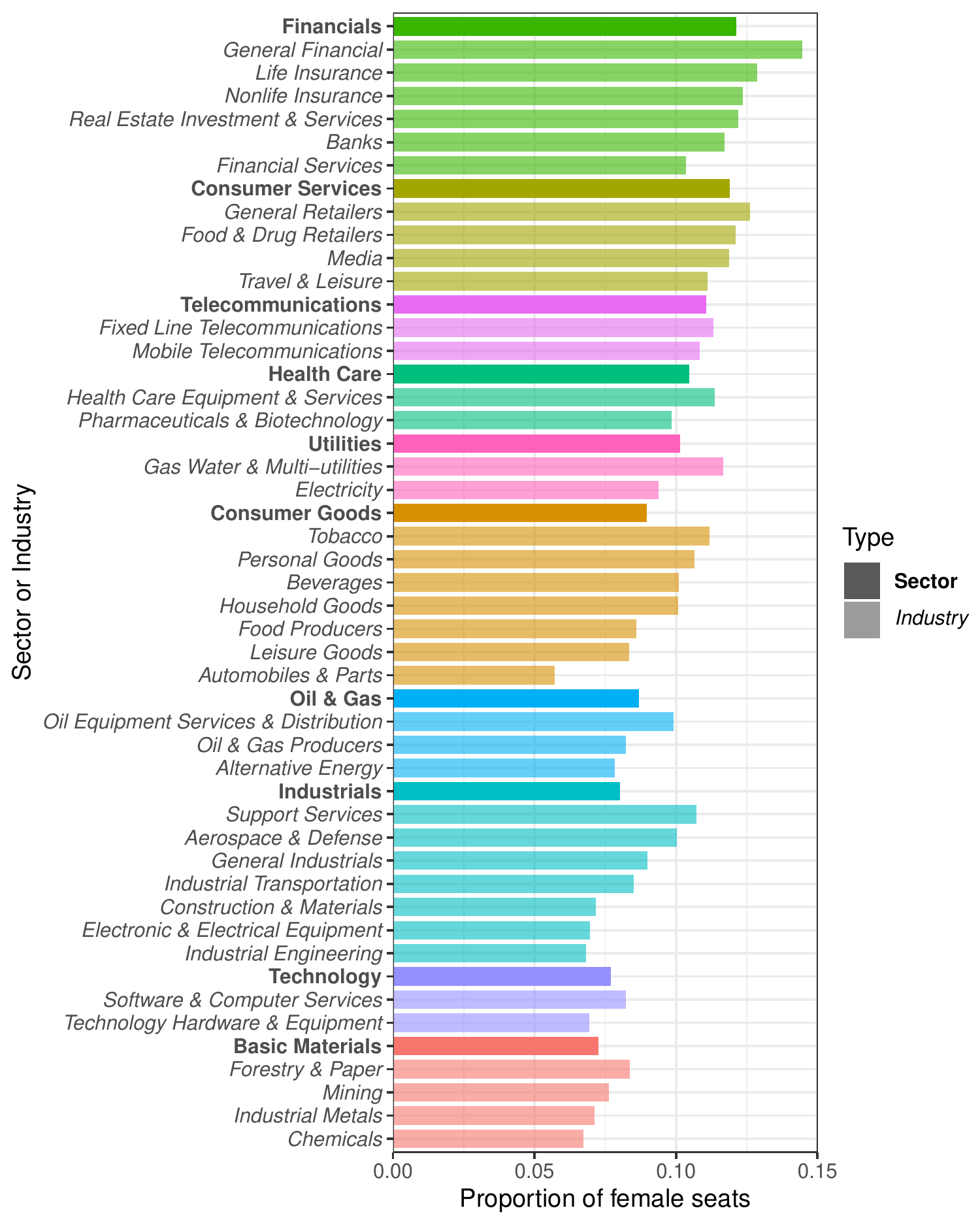}
  \caption{Proportion of female seats by sector (darker colour) and industry (i.e.\ subsector, lighter colour).}
  \label{fig:seats_sec_ind}
\end{figure}

Figure~\ref{fig:seats_sec_ind} plots the proportion by sector and then
subdivides each sector by industry, with industries following their
sectors in descending order. 
We note that women are more represented in Financials, Consumer
Services and Telecommunications than in Technology and Basic
Materials. 
These findings are consistent with the report by Credit
Suisse~\cite{Dawson_etal16}.

\begin{figure}
\centering
\hspace*{-0.5cm}
\includegraphics[height=14cm]{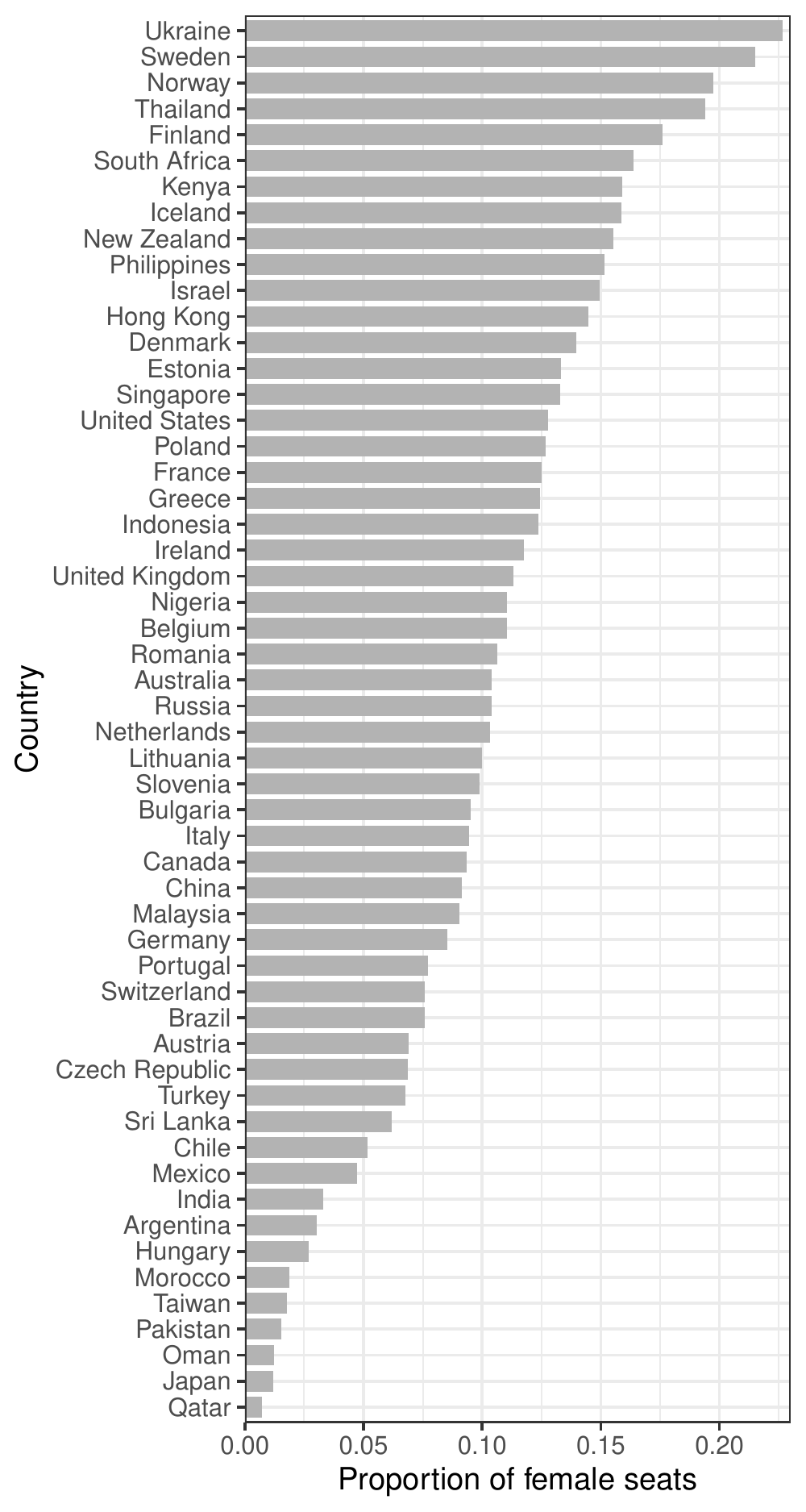}
  \caption{Proportion of female seats by country.}
  \label{fig:seats_country}
\end{figure}

We find greater discrepancy between our data and Credit Suisse when we
split our data by country (Figure~\ref{fig:seats_country}) instead of sector or industry.
While we agree that Scandinavian countries generally rank highly, we find lower percentages of female seats than those reported by both Credit Suisse and Deloitte~\cite{Deloitte17}.
For example, we find the percentage of female seats in Sweden to be
21.52\%, whereas Credit Suisse reports 33.6\% and Deloitte 31.7\%.
We believe that the difference is due to our larger sample size.
The FT data base includes 465 Swedish
companies compared to only 125 in Deloitte's data.

The top-ranked country in our data is Ukraine (22.6\%).
We have not found previous reports on female directors in Ukraine so that we cannot rule out that its top rank is owed to a relatively small sample size of only 19 Ukrainian companies.
Another surprisingly highly ranked country is Thailand (19.5\%). 
Although Deloitte estimates the percentage to be only 11.7\%, it is
plausible that the true number is higher because Thailand is among the
countries with the largest proportion (37\%) of women in senior
management~\cite{GrantThornton16}.
Near the bottom of the ranking is Japan (1.2\%) where our number is
even lower than previous estimates (Credit Suisse: 3.5\%,
Deloitte: 4.1\%).

Similar to the worldwide trend mentioned above, the proportion of female {\it directors} by country hardly differs from the  proportion of female {\it seats} (i.e.\ the data shown in Figure~\ref{fig:seats_country}). We have inferred the country of a person as the most common country of her or his companies. Based on this assumption, we have calculated the countrywide proportion of female directors. In every country included in the FT database, it differs by less than 0.022\% from the proportion of female seats so that the conclusions do not depend on whether we use female seats or female directors as the basis of our analysis.

\begin{figure}
\centering
\includegraphics[height=14cm]{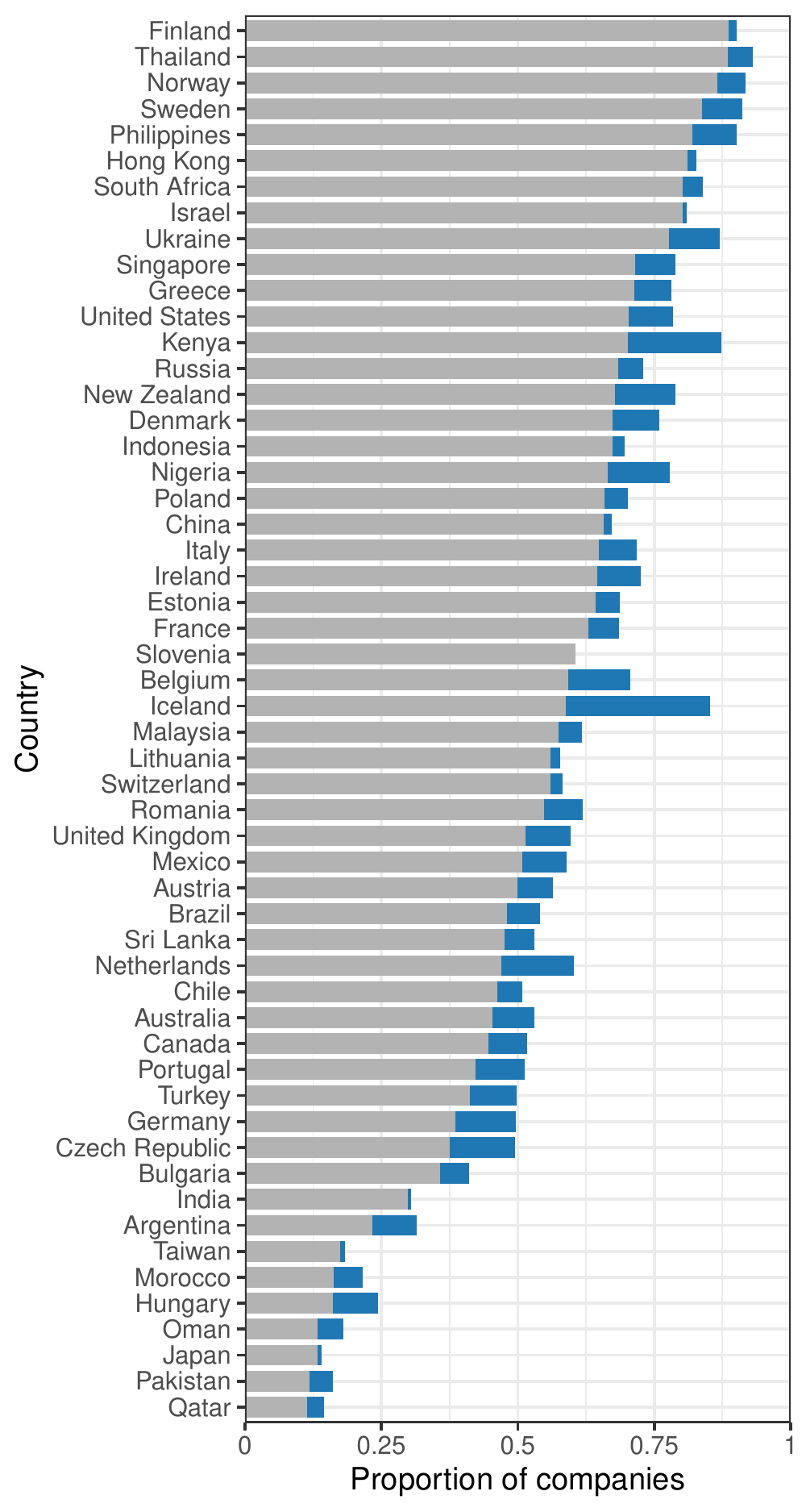}
\caption{Observed and predicted proportions of companies with at least one woman on their boards by country. The observed proportion is shown in grey. The predicted proportion is the combination of the grey and the blue bar. The prediction is calculated under the assumption that seats are taken by both genders independently given the observed proportion of female seats in each country (see text). The prediction is higher than the observed proportion in all cases except Slovenia, where the prediction is only slightly lower.}
\label{pred_wom_comps_by_country}
\end{figure}

Another measure for comparing female representation across countries
is the proportion of companies with at least one woman on their boards. 
Worldwide, we find that 50.3\% of companies have at least one director who FT identifies as female.
Because FT does not include gender information for 30.1\% of the directors (see section~\ref{sec:data}), the true percentage of companies with women on their boards is likely to be higher. Lee et al.~\cite{Lee_etal15} estimate 73.5\%
for the smaller MSCI data base.
The country rankings, shown in Figure~\ref{pred_wom_comps_by_country} (grey bars in the plot), are similar to those for the proportion of female directors by country in Figure~\ref{fig:seats_country}. 
Ukraine drops to the ninth position, but the Scandinavian countries and Thailand maintain their high rankings.
Oman, Japan, Pakistan, and Qatar remain at the bottom.

It is instructive to compare the observed percentage of companies with women on their boards with the expected values from a simple probabilistic null model.
We assume that the probability of a seat being held by a woman is equal to the observed fraction $p$ of female seats in a given country.
In the null model, we assume that the assignment of women to seats is independent of the gender of the other seats.
Suppose the size of a board is $s$.
For each of the $s$ seats, we flip a biased coin which shows heads with probability $p$ and tails with probability $1-p$.
When the coin shows heads, the seat is, in this model, given to a woman.
The probability that the company's board has at least one woman equals $1 - (1-p)^{s}$.
If the fraction of boards with $s$ seats is $f_s$, then the expected fraction of boards with women is $\sum_s f_s [1 - (1-p)^s]$.

The alternative hypothesis is that companies tend to have a single token woman on their boards.
In this hypothesis, a woman is only added when there is currently no other woman on the board~\cite{farrell2005additions,StrydomYong12}.
With exactly one woman, the board satisfies a minimum criterion of diversity that reduces external pressure for greater female representation without seriously threatening the power of the ``old-boys network''.
If the token woman hypothesis is true, there would be a higher proportion of boards with exactly one female board member than in the null model.

We calculated the null model's expectation value for the global data and the predicted proportion of single-woman boards for each country.  Worldwide, we find that the prediction is higher than the observation ($0.604$ vs.\ $0.507$). On the level of individual countries, Slovenia is the only case where the prediction is lower than the observation, but even there the predicted proportion is only lower by $0.0013$. In all other countries, there are fewer single-woman boards than predicted by the null model (Figure~\ref{pred_wom_comps_by_country}). In some cases (e.g.\ Iceland or the United States) the difference is substantial. This implies that women are generally more clustered than expected if they were distributed randomly, contradicting the token woman hypothesis.

\begin{figure}
  \centering
  \includegraphics[width=0.9\textwidth]{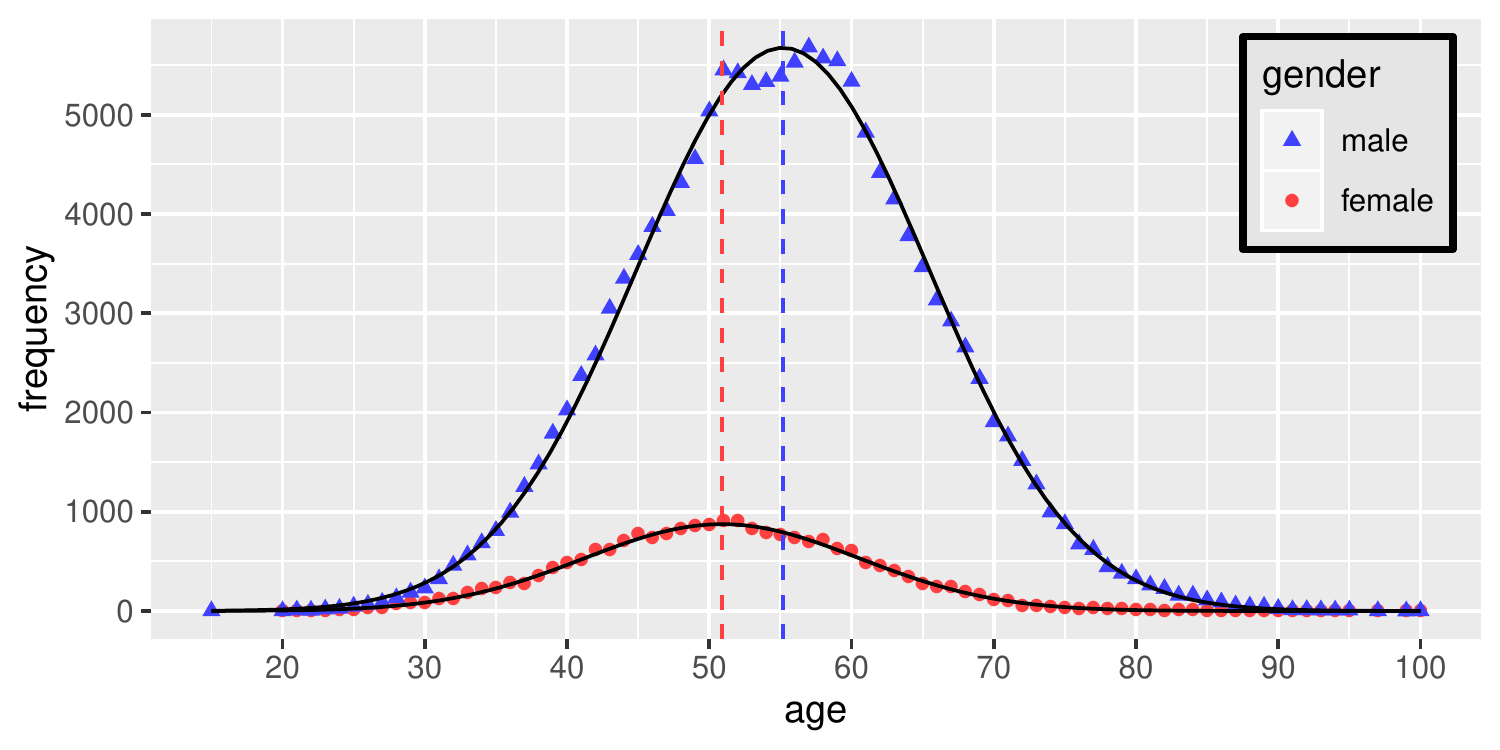}
  \caption{The age distribution of directors by gender. Both distributions are well approximated by Gaussian functions (black solid curves), but with different means (dashed lines): the mean age of male directors is $55.1$ years, that of female directors $50.8$ years.
  The FT data contain information about the age of more than 200\,000 directors.
  Note that 164 directors had their age listed as 1 year. We have removed these data points from our analysis.}
  \label{age_dist}
\end{figure}

Although we find no evidence that women are recruited as tokens, their number has increased in recent years.
As a consequence, women directors are, on average, younger than their male colleagues.
In Figure~\ref{age_dist}, we plot the age distribution by gender for all directors for whom the FT database contains information about both age and gender. 
The figure makes it clear that the distributions are centered at a different age: the mean is 50.8 years for women and 55.1 for men.

The solid curves in Figure~\ref{age_dist} show that the age distribution is approximately normal for both genders.
Strictly speaking, neither the female nor the male distribution passes a $\chi^2$-test for normality ($p$-values $<10^{-9}$) because they are slightly skewed towards higher age.
However, the deviations from normality are sufficiently small to justify using Welch's two-sample $t$-test.
The null hypothesis of an equal mean for men and women is strongly rejected ($p$-value $< 10^{-15}$).
Our result is consistent with earlier studies of French~\cite{Lee_etal15} and Singaporean data~\cite{SG-DAC16} where female directors were found to be on average 5--10 years younger.

\section{The position of women in the network}

While the attributes discussed in the previous section already give us some insight into gender differences, we can only truly assess the role of women when considering their positions in the network.
As we explained in the introduction, our data can be viewed as a bipartite network where edges run between directors and boards (Figure~\ref{fig:netw}a).
In this network, there are 2\,809\,623 edges connecting 321\,967 directors to 35\,927 boards.
The average board size (of those available) is 11.02.
The mean size of a board without women is 8.88, whereas boards with at least one woman have on average 13.10 seats.
Some care needs to be taken when interpreting these numbers. 
Even in our earlier null model, where we assumed that seats are independently taken by men and women, the mean size of a board with a woman is larger than the mean size without a woman.
The reason is that, in this model, the probability of at least one woman on a board of size $s$ is $1-(1-p)^s$ and thus increases with $s$.
The mean board size conditioned on the presence of at least one woman is
\[
\mu_\textnormal{null} \equiv
E[\textnormal{board size} \mid \textnormal{woman}] = \frac{\sum_{s=1}^\infty s f_s [1-(1-p)^s]} {\sum_{s=1}^\infty f_s [1-(1-p)^s]}\ ,
\]
where $f_s$ is, as before, the fraction of boards with $s$ seats.
We find $\mu_\textnormal{null} = 12.97$, comparable to the observed value 13.10, but statistically significantly smaller ($p$-value $<10^{-8}$). 
This result confirms previous observations that larger boards tend to have a higher probability of recruiting women~\cite{Burke00,Carter_etal03,NguyenFaff06}.

\begin{table}
\renewcommand{\arraystretch}{1.2}
\renewcommand{\aboverulesep}{0pt}
\renewcommand{\belowrulesep}{0pt}
\centering
  \caption{Summary statistics of the network with all nodes (i.e.\ nodes identified as male, female, and those with missing gender information) and all edges. Larger values are highlighted in bold. ``Like degree'' is the degree between nodes of the same gender.
  We calculate the closeness centrality with the harmonic mean formula by Newman~\cite{Newman10} to handle disconnected components.}
  \label{table:all_nodes}
  \begin{minipage}{\textwidth}
    \begin{tabular}{p{0.45\textwidth}p{0.17\textwidth}p{0.17\textwidth}p{0.16\textwidth}}
      \toprule
    \rowcolor{LightGrey} \textbf{nodes} & \textbf{all} & \textbf{male} & \textbf{female}\\
      \midrule
 \% of all & 100 & \textbf{60.5} & 9.43 \\ 
\hline
\% in the largest component & 74.7 & 74.8 & \textbf{79.4} \\ 
\hline
maximum \textbf{degree} & 1040 & 1016 & \textbf{1030} \\ 
\hline
average \textbf{degree} & 17.45 & 16.68 & \textbf{18.74} \\ 
\hline
average \textbf{``like degree''} & & \textbf{11.21} & 2.94 \\ 
\hline
average \textbf{degree} & \multirow{2}{*}{19.97} & \multirow{2}{*}{19.05} & \multirow{2}{*}{\textbf{20.98}} \\
in largest component\\
\hline
maximum \textbf{betweenness centrality} & $9.864\times 10^{8}$ & $\mathbf{9.864\times 10^{8}}$ & $4.755\times 10^{8}$ \\ 
\hline
average \textbf{betweenness centrality} & $6.515\times 10^{5}$ & $6.556\times 10^{5}$ & $\mathbf{8.519\times 10^{5}}$ \\ 
\hline
average \textbf{betweenness centrality} & \multirow{2}{*}{$8.709\times 10^{5}$} & \multirow{2}{*}{$8.753\times 10^{5}$} & \multirow{2}{*}{$\mathbf{10.72\times 10^{5}}$} \\
in largest component\\
\hline
maximum \textbf{closeness centrality} & 0.143 & 0.143  & 0.143 \\ 
\hline
average \textbf{closeness centrality} & $0.071$ & $0.072$ & \textbf{0.078} \\ 
\hline
average \textbf{closeness centrality} & \multirow{2}{*}{0.096} & \multirow{2}{*}{\textbf{0.099}} & \multirow{2}{*}{0.097} \\
in largest component\\
\hline
maximum \textbf{clustering coefficient} & 1 & 1 & 1 \\
      \hline
average \textbf{clustering coefficient} & 0.939 & 0.935 & \textbf{0.936}\\
      \hline
average \textbf{clustering coefficient} & \multirow{2}{*}{0.9289} & \multirow{2}{*}{0.9243} & \multirow{2}{*}{\textbf{0.9272}} \\
in largest component\\
\bottomrule
    \end{tabular}
    \vspace{-\baselineskip}
  \end{minipage}
  \label{table_all}
\end{table}


Larger boards tend to be in the largest component of the bipartite network.
In the one-mode projection that only contains the directors as nodes (Figure~\ref{fig:netw}b), the largest component
consists of 74.7\% of the nodes and 85.5\% of the edges.
Given that women are more likely to be on larger boards, it is not surprising that the proportion of women in the largest component (79.4\%) exceeds the fraction of nodes belonging to that component.
We confirm with the $\chi^2$-test proposed by Hawarden \& Marsland~\cite{HawardenMarsland11} that the proportion of women in the largest component is significantly higher than that of men ($p$-value $<10^{-15})$.
We therefore agree with their previous result that, although women are a minority, they are not marginalised by being confined to unconnected, and hence less influential, components.

In terms of degree and betweenness centrality statistics, women are doing marginally better than men (Table~\ref{table_all}). The distributions of degree and betweenness centrality by gender are not normal but instead seem to follow power laws. We normalise them by log-transforming the data and restricting our sample to the largest component and nodes with the parameter of interest $>0$. The two-sample $t$-test for degree concludes that the marginal difference between men and women is statistically significant (p-value $<0.0001$). The difference in the betweenness centrality is not statistically significant at a significance level of 0.05 ($p$-value 0.068).

\section{Conclusion}
In this paper we have analysed a new dataset which allows us to better understand interlocking directorates. In particular, it has allowed us to show the differences of female representation by country and industry. Overall, there are still many fewer women than men on boards, but our analysis contradicts the token woman hypothesis whereby companies recruit exactly one woman to escape accusations of discrimination.
A limitation of our dataset is that it only indicates presence or absence of a link, but not its strength, which has been hypothesised to depend on gender~\cite{ibarra1992homophily,Ibarra93}.
Our binary data, however, do not show evidence that women are marginalised.

\medskip
\noindent
\textbf{Acknowledgements:}
We would like to thank Adrian Roellin for introducing us to the study
of interlocks and to the Financial Times Equities database.
M.~T.~G. was supported by the Singapore Ministry of Education and a
Yale-NUS College start-up grant (R-607-263-043-121).

%
%
\bibliography{women_on_board}

\begin{thebibliography}{10}
\providecommand{\url}[1]{\texttt{#1}}
\providecommand{\urlprefix}{URL }

\bibitem{adams2016women}
Adams, R.B.: Women on boards: the superheroes of tomorrow? Leadersh.~Q.  27(3),
   371--386 (2016)

\bibitem{BattistonCatanzaro04}
Battiston, S., Catanzaro, M.: Statistical properties of corporate board and
  director networks. Eur. Phys. J. B  38(2),  345--352 (2004)

\bibitem{BurgessTharenou02}
Burgess, Z., Tharenou, P.: Women board directors: characteristics of the few.
  J.~Bus.~Ethics  37(1),  39--49 (2002)

\bibitem{Burke00}
Burke, R.J.: Company Size, Board Size and Numbers of Women Corporate Directors,
  pp. 157--167. Springer Netherlands, Dordrecht (2000)

\bibitem{Carter_etal03}
Carter, D.A., Simkins, B.J., Simpson, W.G.: Corporate governance, board
  diversity, and firm value. Financ. Rev.  38(1),  33--53 (2003)

\bibitem{Dawson_etal16}
Dawson, J., Kersley, R., Natella, S.: The {CS} {G}ender 3000: the reward for
  change. Tech. rep., Credit Suisse Research Institute (2016)

\bibitem{Dawson_etal16b}
Dawson, J., Kersley, R., Vair, B., Preto, M.: Overboarding in the {US}. Tech.
  rep., Credit Suisse ESG Research (2016)

\bibitem{delis2016effect}
Delis, M.D., Gaganis, C., Hasan, I., Pasiouras, F.: The effect of board
  directors from countries with different genetic diversity levels on corporate
  performance. Manag. Sci.  63(1),  231--249 (2016)

\bibitem{Deloitte17}
Deloitte: Women in the boardroom: a global perspective. Tech. rep., Deloitte
  Global Center for Corporate Governance (2017)

\bibitem{dezsHo2016there}
Dezs{\H{o}}, C.L., Ross, D.G., Uribe, J.: Is there an implicit quota on women
  in top management? {A} large-sample statistical analysis. Strategic Manag. J.
   37(1),  98--115 (2016)

\bibitem{dHoop_etal17}
{d'Hoop-Azar}, A., Martens, K., Papolis, P., Sancho, E.: Gender parity on
  boards around the world. Harvard Law School Forum on Corporate Governance and
  Financial Regulation (2017), retrieved from
  https://corpgov.law.harvard.edu/2017/01/05/gender-parity-on-boards-around-the-world/\#more-76896
  on 21 May 2017

\bibitem{Eagly16}
Eagly, A.H.: When passionate advocates meet research on diversity, does the
  honest broker stand a chance? J. Soc. Issues  72(1),  199--222 (2016)

\bibitem{elango2018women}
Elango, B.: When do women reach the top spot? {A} multilevel study of female
  {CEO}s in emerging markets. Manag. Decis.  (2018), in press

\bibitem{EuropeanCommission15}
{European Commission}: Gender balance on corporate boards: Europe is cracking
  the glass ceiling (2015), retrieved from
  http://ec.europa.eu/justice/gender-equality/files/womenonboards/factsheet\_women\_on\_boards\_web\_2015-10\_en.pdf
  on 23 May 2017

\bibitem{farrell2005additions}
Farrell, K.A., Hersch, P.L.: Additions to corporate boards: the effect of
  gender. J.~Corp.~Finance  11(1-2),  85--106 (2005)

\bibitem{gabaldon2016searching}
Gabaldon, P., De~Anca, C., Mateos~de Cabo, R., Gimeno, R.: Searching for women
  on boards: an analysis from the supply and demand perspective. Corp.~Gov.
  24(3),  371--385 (2016)

\bibitem{GrantThornton16}
{Grant Thornton}: Women in business: turning promise into practice (2016),
  \\retrieved from
  https://www.grantthornton.global/globalassets/wib\_turning\_promise\_into\_practice.pdf
  on 20 May 2017

\bibitem{Hawarden10}
Hawarden, R.J.: Women on boards of directors: the origin and structure of
  gendered small-world and scale-free director glass networks. Ph.D. thesis,
  Massey University, Palmerston North (2010)

\bibitem{HawardenMarsland11}
Hawarden, R.J., Marsland, S.: Locating women board members in gendered director
  networks. Gend. Manag.  26(8),  532--549 (2011)

\bibitem{Hillman_etal07}
Hillman, A.J., Shropshire, C., Cannella, A.A.: Organizational predictors of
  women on corporate boards. Acad. Manag. J.  50(4),  941--952 (2007)

\bibitem{ibarra1992homophily}
Ibarra, H.: Homophily and differential returns: Sex differences in network
  structure and access in an advertising firm. Adm. Sci. Q.  37(3),  422--447
  (1992)

\bibitem{Ibarra93}
Ibarra, H.: Personal networks of women and minorities in management: a
  conceptual framework. Acad. Manag. Rev  18(1),  56--87 (1993)

\bibitem{Izraeli00}
Izraeli, D.: The paradox of affirmative action for women directors in {I}srael.
  In: Burke, R.J., Mattis, M.C. (eds.) Women on corporate boards of directors:
  international challenges and opportunities, pp. 75--96. Springer Netherlands,
  Dordrecht (2000)

\bibitem{Lee_etal15}
Lee, L.E., Marshall, R., Rallis, D., Moscardi, M.: Women on boards (2015),
  retrieved from
  https://www.msci.com/documents/10199/04b6f646-d638-4878-9c61-4eb91748a82b on
  26 July 2017

\bibitem{LevineRoy79}
Levine, J.H., Roy, W.S.: A study of interlocking directorates: vital concepts
  of organization. In: Holland, P.W., Leinhardt, S. (eds.) Perspectives on
  {S}ocial {N}etwork {R}esearch, pp. 349--378. Academic Press, New York (1979)

\bibitem{Martinez12}
Mart\'inez, A.C.: Social Network Analysis and the illusion of gender neutral
  organisations. Master's thesis, Universitat Polit\`ecnica de Catalunya (2012)

\bibitem{mcgregor_2017}
McGregor, J.: The number of women {CEO}s in the {F}ortune 500 is at an all-time
  high - of 32. The Washington Post (7 Jun 2017)

\bibitem{Newman10}
Newman, M.: Networks: an introduction. Oxford University Press, Inc., New York,
  NY, USA (2010)

\bibitem{NguyenFaff06}
Nguyen, H., Faff, R.: Impact of board size and board diversity on firm value:
  {A}ustralian evidence. Corp. Ownersh. Control  4(2),  24--32 (2006)

\bibitem{Perrault15}
Perrault, E.: Why does board gender diversity matter and how do we get there?
  {T}he role of shareholder activism in deinstitutionalizing old boys'
  networks. J.~Bus.~Ethics  128(1),  149--165 (Apr 2015)

\bibitem{post2015women}
Post, C., Byron, K.: Women on boards and firm financial performance: a
  meta-analysis. Acad.~Manag.~J.  58(5),  1546--1571 (2015)

\bibitem{SeierstadOpsahl11}
Seierstad, C., Opsahl, T.: For the few not the many? {T}he effects of
  affirmative action on presence, prominence, and social capital of women
  directors in {N}orway. Scand.~J.~Manag.  27(1),  44--54 (2011)

\bibitem{sojo2016reporting}
Sojo, V.E., Wood, R.E., Wood, S.A., Wheeler, M.A.: Reporting requirements,
  targets, and quotas for women in leadership. Leadersh. Q.  27(3),  519--536
  (2016)

\bibitem{SonquistKoenig75}
Sonquist, J.A., Koenig, T.: Interlocking directorates in the top {U.S.}
  corporations. Insurgent {S}ociologist  5(3),  196--229 (1975)

\bibitem{StrydomYong12}
Strydom, M., Yong, H.H.A.: The token woman. In: 25th {A}ustralasian {F}inance
  and {B}anking {C}onference (2012), available at
  http://dx.doi.org/10.2139/ssrn.2136737

\bibitem{WorldBank17}
{The World Bank}: Labor force, female (\% of total labor force) (2017),
  retrieved from http://data.worldbank.org/indicator/SL.TLF.TOTL.FE.ZS on 20
  May 2017

\bibitem{ThomsonGraham05}
Thomson, P., Graham, J.: A Woman’s Place is in the Boardroom. Palgrave
  Macmillan, Basingstoke (2005)

\bibitem{ThomsonReuters16}
{Thomson Reuters Corporation}: Profiles and lists of directors of publicly
  traded companies. https://markets.ft.com/data/equities/results (2016),
  retrieved on 17 September 2016

\bibitem{WestphalMilton00}
Westphal, J.D., Milton, L.P.: How experience and network ties affect the
  influence of demographic minorities on corporate boards. Adm. Sci. Q.  45(2),
   366--398 (2000)

\bibitem{SG-DAC16}
Yacob, H., et~al.: Women on boards: tackling the issue. Tech. rep., Singapore's
  Diversity Action Committee (2016)

\end{thebibliography}

\end{document}